\documentclass[journal=jctcce,manuscript=article]{achemso}
\usepackage[version=3]{mhchem} 
\usepackage{caption}
\usepackage{subcaption}
\usepackage{hyperref}
\usepackage{natbib}
\usepackage{siunitx}
\usepackage{tabularx}
\usepackage{amsmath}
\usepackage{xcolor}
\usepackage{threeparttable}
\usepackage{multirow}

\newcommand{\Iaq}{{I\alpha,\mathbf{q}}}
\newcommand{\Jbq}{{J\beta,\mathbf{q}}}
\newcommand{\bfk}{{\mathbf{k}}}
\newcommand{\bfq}{{\mathbf{q}}}

\author{Han Yang}
\affiliation{Department of Chemistry, University of Chicago, Chicago, Illinois 60637, United States}
\alsoaffiliation{Pritzker School of Molecular Engineering, The University of Chicago, Chicago, Illinois 60637, United States}
\author{Marco Govoni}
\affiliation{Materials Science Division and Center for Molecular Engineering, Argonne National Laboratory, Lemont, Illinois 60439, United States}
\alsoaffiliation{Pritzker School of Molecular Engineering, The University of Chicago, Chicago, Illinois 60637, United States}
\email{mgovoni@anl.gov}
\author{Arpan Kundu}
\affiliation{Pritzker School of Molecular Engineering, The University of Chicago, Chicago, Illinois 60637, United States}
\author{Giulia Galli}
\affiliation{Department of Chemistry, University of Chicago, Chicago, Illinois 60637, United States}
\alsoaffiliation{Pritzker School of Molecular Engineering, The University of Chicago, Chicago, Illinois 60637, United States}
\alsoaffiliation{Materials Science Division and Center for Molecular Engineering, Argonne National Laboratory, Lemont, Illinois 60439, United States}
\email{gagalli@uchicago.edu}

\title{Combined first-principles calculations of electron-electron and electron-phonon self-energies in condensed systems}
\begin{document}
\maketitle
\begin{abstract}
    We present a method to efficiently combine the computation of electron-electron and electron-phonon self-energies, which enables the evaluation of electron-phonon coupling at the $G_0W_0$ level of theory for systems with hundreds of atoms. In addition, our approach, which is a generalization of a method recently proposed for molecules [\textit{J. Chem. Theory Comput.} 2018, \textbf{14}, 6269–6275], enables the inclusion of non-adiabatic and temperature effects at no additional computational cost. We present results for diamond and defects in diamond and discuss the importance of numerically accurate $G_0W_0$ band structures to obtain robust predictions of zero point renormalization (ZPR) of band gaps, and of the inclusion of non-adiabatic effect to accurately compute the ZPR of defect states in the band gap.  
\end{abstract}
\section{Introduction}
The interaction between electrons and phonons in solids\cite{giustino2017electron,giustino2010electron,antonius2014many}  gives rise to a variety of interesting physical phenomena, including superconductivity\cite{bardeen1957microscopic},  and to complex electronic structure properties in metals, semiconductors and insulators\cite{fan1951temperature}.
Electron-phonon coupling has been widely studied for more than half a century\cite{frohlich1950xx}. However, it is only in the last two decades that first principles, quantum mechanical methods have been applied to carry out quantitative calculations\cite{giustino2007EPW,giustino2017electron,antonius2014many}, based on  frozen-phonon approaches\cite{monserrat2014,monserrat2016correlation,monserrat2018electron,antonius2014many,karsai2018electron}, density functional perturbation theory (DFPT)\cite{giannozzi1991ab,baroni2001phonons,giustino2017electron,antonius2014many,ponce2014temperature,cannuccia2011effect,cannuccia2012zero} and, very recently, path-integral molecular dynamics simulations based on density functional theory (DFT)\cite{kundu2021quantum}. 

The frozen-phonon approach is straightforward to implement, compared to other methods, as the phonon frequencies and electron-phonon renormalization energies are simply computed by displacing the nuclear positions and solving the Kohn-Sham equations at each displaced position\cite{antonius2014many,capaz2005frozenphonon,monserrat2014,monserrat2016correlation}. However, the frozen-phonon approach is difficult to converge with respect to the supercell size, \cite{baroni2001phonons,giustino2017electron} and using this method it is challenging to accurately describe polar systems\cite{ponce2015temperature_JCP,miglio2020nonadiabatic}. Hence, perturbative approaches have been widely used. Most of them  solve the electronic structure problem at the level of density functional theory (DFT)\cite{antonius2014many,cannuccia2011effect,cannuccia2012zero} and compute electron-phonon coupling within the Allen-Heine-Cardona (AHC) formalism\cite{allen1976theory,allen1981theory,giustino2010electron}. However, some formulations have been recently proposed to  go beyond the approximations of the AHC approach, and  include non-adiabatic
\cite{ponce2014temperature,ponce2015temperature_JCP,miglio2020nonadiabatic}, dynamical\cite{cannuccia2011effect,cannuccia2012zero,antonius2015dynamical} and/or non-rigid-ion\cite{ponce2014temperature} effects in the calculation of electron-phonon interactions. It has also been shown that for several solids many-body perturbation theory (MBPT)\cite{giustino2010electron,antonius2014many,monserrat2016correlation} is necessary, in order to obtain  results in quantitative agreement with experiments, and  $GW$ corrections have been applied to compute electron-coupling matrices\cite{li2019elph_berkeleygw,li2021unmasking} and/or DFT single particle energy levels\cite{giustino2010electron,antonius2014many,mcavoy2018coupling}. Given the computational cost involved in electron-phonon calculations based on MBPT, e.g., at the $G_0W_0$ level, plasmon-pole models (PPM) are often employed\cite{ppm1986,antonius2014many,monserrat2016correlation,li2019elph_berkeleygw,godby1988self} to approximate the frequency dependence of the self-energy, in spite of some known deficiencies of such models\cite{martin2016interacting}.   

With the goal of improving the accuracy and efficiency of electron-phonon calculations within MBPT,  we recently proposed \cite{mcavoy2018coupling} a method that  combines the evaluation  of electron–electron and electron–phonon self-energies. The dielectric matrix is represented in terms of dielectric eigenpotentials\cite{wilson2008efficient,wilson2009iterative}, utilized for both the calculation of $G_0W_0$ quasi-particle energies and the diagonalization of the dynamical matrix; virtual electronic states are not explicitly computed, dielectric matrices, being  represented using a spectral decomposition, are never inverted, and all self-energies are evaluated over the full frequency spectrum using the Lanczos algorithm.\cite{pham2013g,govoni2015large,govoni2018gw100}. Importantly, our implementation also enables at no extra cost the evaluation of non-adiabatic effects and electron-phonon self energies at multiple temperatures and frequencies. Although in principle the method is general, in practice it  has so far been applied only to finite systems within the adiabatic approximation.

In this work, we generalize the combined electron-electron and electron-phonon approach described in  Ref.~\citenum{mcavoy2018coupling} to solids and, after presenting a detailed verification and validation protocol, we apply the approach to large supercells with about 1000 electrons, which are representative of defective solids. We report results as a function of temperature and we study in detail the effect of including non-adiabatic terms in the zero point renormalization of the band gap of pristine and defective diamond and on defect states.

The rest of this work is organized as follows: we first describe our methodology and then verify our implementation and validate our results for specific systems; we then present applications of the method to  point defects in diamond.


\section{Methodology}\label{sec:method}
\subsection{Dynamical and electron-phonon coupling matrices}
In a periodic system, phonon frequencies are computed by diagonalizing the dynamical matrix\cite{baroni2001phonons}
\begin{equation}
    D_{I\alpha,J\beta}(\mathbf{q}) = \frac{1}{\sqrt{M_IM_J}}C_{I\alpha,J\beta}(\mathbf{q}) = \frac{1}{\sqrt{M_IM_J}}\frac{\partial^2E}{\partial u_{I\alpha}^*(\mathbf{q})\partial u_{J\beta}(\mathbf{q})}
\end{equation}
where $C_{I\alpha,J\beta}(\mathbf{q})$ is a force constant, $E$ is the total energy of the system, $u$ denotes displacements from equilibrium atomic positions, $M_I$, $M_J$ are atomic masses, $I$, $J$ are indices of atoms, $\alpha$, $\beta$ are Cartesian directions, and $\mathbf{q}$ is the wave-vector of the phonon mode.

The force constants are given by the sum of an  electronic and ionic part, with the latter being trivial to evaluate. Within the framework of density functional perturbation theory (DFPT), the electronic contribution can be written as
\begin{equation}
    C^{elec}_{I\alpha,J\beta}(\mathbf{q}) = \sum_{n}^{occ}\sum_\mathbf{k}\left\langle
    \psi_{n\mathbf{k}}
    \left|
    \partial_{J\beta,\mathbf{q}}V_{ext}
    \right|
    \partial_{I\alpha,\mathbf{q}}\psi_{n\mathbf{k}}
    \right\rangle  +c.c. ,\label{equ:fconst}
\end{equation}
where $\mathbf{k}$ is a k-point within the Brillouin zone, $\psi_{n\mathbf{k}}$ is the wavefunction of the $n$-th band at $\mathbf{k}$, and $V_{ext}$ is the external ionic potential. For simplicity, we denoted the derivative $\partial/\partial u_{I\alpha}(\mathbf{q})$ as $\partial_{I\alpha,\mathbf{q}}$. The braket in Eq.~\ref{equ:fconst} is commonly computed by solving the Sternheimer equation self-consistently,
\begin{equation}
    (\hat{H}_{KS}-\varepsilon_{n\mathbf{k}})\left|\partial_{I\alpha,\mathbf{q}}\psi_{n\mathbf{k}}\right\rangle = -\hat{\mathcal{P}}^c_{\mathbf{k}+\mathbf{q}}\partial_{I\alpha,\mathbf{q}} \hat{V}_{SCF}\left|\psi_{n\mathbf{k}}\right\rangle,
\end{equation}
where $\hat{H}_{KS} = \hat{K} + \hat{V}_{SCF}$ is the Kohn-Sham Hamiltonian; $\hat{K}$ is the kinetic operator; $\hat{V}_{SCF}$ is the self-consistent potential operator; $\varepsilon_{n\mathbf{k}}$ is the Kohn-Sham eigenvalue of the $n$-th band at the $\mathbf{k}$ point, $\partial_{I\alpha,\mathbf{q}} \hat{V}_{SCF}$ is the first order change of the self-consistent potential due to atomic displacements,  $\partial_{I\alpha,\mathbf{q}} \psi_{n\mathbf{k}}$ denotes the first order change of the wavefunction, and $\hat{\mathcal{P}}^c_{\mathbf{k}+\mathbf{q}} = \hat{I}-\sum_{v}^{occ}\left|\psi_{v\mathbf{k}+\mathbf{q}}\right\rangle\left\langle\psi_{v\mathbf{k}+\mathbf{q}}\right|$ is the projection operator onto the  manifold of unoccupied (virtual) single particle electronic states.

Instead of solving the Sternheimer equation self-consistently, we write the braket in Eq.~\ref{equ:fconst} as:\cite{mcavoy2018coupling}
\begin{equation}
    \left\langle \psi_{n\mathbf{k}}
    \left|
    \partial_\Jbq V_{ext}
    \right|
    \partial_{I\alpha,\mathbf{q}}\psi_{n\mathbf{k}}
    \right\rangle
    = 
    \left\langle
    \partial_\Jbq \psi_{n\mathbf{k}}^{bare}
    \left|
    \partial_\Iaq V_{SCF}
    \right|
    \psi_{n\mathbf{k}}
    \right\rangle,
\end{equation}
where the change of the wavefunction is obtained through the one-shot solution of the Sternheimer equation
\begin{equation}
    (\hat{H}_{KS}-\varepsilon_{n\mathbf{k}})\left|\partial_\Iaq \psi_{n\mathbf{k}}^{bare}\right\rangle = -\hat{\mathcal{P}}^c_{\mathbf{k}+\mathbf{q}}\partial_\Iaq \hat{V}_{ext}\left|\psi_{n\mathbf{k}}\right\rangle.
\end{equation}
The change of SCF potential can be evaluated from $\partial_\Iaq \hat{V}_{ext}$ as
\begin{equation}
    \partial_\Iaq V_{SCF} = \partial_\Iaq V_{ext} + [f_{Hxc}+f_{Hxc}\chi f_{Hxc}]\partial_\Iaq \rho^{bare} \label{equ:dVscf}
\end{equation}
where $f_{Hxc} = v_c +f_{xc}$ is the sum of the bare Coulomb potential, $v_c$, and the exchange-correlation kernel, $f_{xc}$; $\chi$ is the reducible density-density response function and $\partial_\Iaq\rho^{bare}$ is the derivative of the bare change of density,
\begin{equation}
    \partial_\Iaq \rho^{bare} = \sum_{n}^{occ}\sum_\mathbf{k} \left[\psi_{n\bfk}^*\partial_\Iaq\psi_{n\bfk}^{bare} + c.c. \right]
\end{equation}

In order to efficiently evaluate the reducible density-density response function, $\chi$, we represent the irreducible density-density response function, $\chi_0$, in terms of projective dielectric eigenpotentials (PDEP)\cite{wilson2009iterative,pham2013g,govoni2015large,yang2019improving} and we represent $\chi$ with the same basis used as that of $\chi_0$:
\begin{equation}
    \chi_0(\bfq) = \sum_i^{N_\mathrm{PDEP}} \left|\phi_{i}(\bfq)\right\rangle\lambda_i(\bfq)\left\langle\phi_i(\bfq)\right|,
\end{equation}
where $i$ is the index of the PDEP basis, $\phi_i(\bfq)$ and $\lambda_i(\bfq)$ are the $i$-th eigenvector and eigenvalue of the symmetrized reducible polarizability, $N_\mathrm{PDEP}$ is the number of PDEP basis functions, respectively. Using the Dyson equation $\chi = \chi_0 +\chi_0 f_{Hxc} \chi$, the reducible density-density response function can be evaluated using the PDEP basis set:
\begin{equation}
    \chi = (1-\chi_0 f_{Hxc})^{-1}\chi_0.
\end{equation}

Adopting the procedure described above, we can compute the dynamical matrix $D(\bfq)$:
\begin{equation}
    \sum_{J\beta}D_{I\alpha,J\beta}(\bfq) \xi_{J\beta,\bfq\nu} = \omega^2_{\bfq\nu}\xi_{I\alpha,\bfq\nu},
    \label{eq:phonon}
\end{equation}
where $\omega_{\bfq\nu}$ are phonon frequencies and $\xi_{I\alpha,\bfq\nu}$ are phonon eigenvectors. Finally, Eq.~\ref{equ:dVscf} is used to evaluate the electron-phonon coupling matrix elements $g$ given by:\cite{giustino2017electron}
\begin{equation}
    g_{mn\nu}(\bfk,\bfq) = \left\langle \psi_{m\bfk+\bfq} \left| \partial_{\bfq\nu}V_{SCF} \right| \psi_{n\bfk} \right\rangle,
    \label{eq:g}
\end{equation}
where $\partial_{\bfq\nu}V_{SCF}$ is the mode-resolved change of potential,
\begin{equation}
\partial_{\bfq\nu}V_{SCF} = \sum_{I\alpha}\frac{\xi_{I\alpha,\bfq\nu}}{\sqrt{2M_I\omega_{\bfq\nu}}}\partial_{I\alpha,\bfq}V_{SCF}
\end{equation}

\subsection{Electron-phonon self-energy}
Within many-body perturbation theory (MBPT)\cite{giustino2017electron,martin2016interacting}, the electron-phonon self-energy has two components, the Fan-Migdal:
\begin{equation}
    \Sigma_{n\mathbf{k}}^\mathrm{FM}(\omega,T) = \sum_{m\nu\mathbf{q}}\left|g_{mn\nu}(\mathbf{k},\mathbf{q})\right|^2\left[\frac{n_{\mathbf{q}\nu}+f_{m\mathbf{k}+\mathbf{q}}}{\omega-\varepsilon_{m\mathbf{k}+\mathbf{q}}+\omega_{\mathbf{q}\nu}-i0^+}+\frac{n_{\mathbf{q}\nu}+1-f_{m\mathbf{k}+\mathbf{q}}}{\omega-\varepsilon_{m\mathbf{k}+\mathbf{q}}-\omega_{\mathbf{q}\nu}-i0^+}\right]
    \label{eq:FM}
\end{equation}
and Debye-Waller:
\begin{equation}
    \Sigma_{n\mathbf{k}}^{DW}(T)=-\sum_{m\nu\mathbf{q}}\sum_{I\alpha J\beta}\frac{2n_{\mathbf{q}\nu}+1}{\varepsilon_{n\mathbf{k}}-\varepsilon_{m\mathbf{k}}}\frac{1}{4\omega_{\mathbf{q}\nu}}\left[\frac{\xi_{I\alpha,\mathbf{q}\nu}\xi_{I\beta,\mathbf{q}\nu}^*}{M_I} +\frac{\xi_{J\alpha,\mathbf{q}\nu}\xi_{J\beta,\mathbf{q}\nu}^*}{M_J}\right]g^{*,I\alpha}_{mn}(\mathbf{k},\mathbf{0})g^{J\beta}_{mn}(\mathbf{k},\mathbf{0}),
    \label{eq:DW}
\end{equation}
where $n_{\bfq\nu}$ and $f_{m\bfk+\bfq}$ are Bose-Einstein and Fermi-Dirac distributions, respectively. We note that the expression of the Debye-Waller self-energy of Eq. 14 is written by  assuming the rigid-ion approximation, in which the second-order expression of the electron-phonon coupling matrix elements are approximated with their respective  first-order expressions. The effect of this approximation has been thoroughly studied in Ref. \citenum{ponce2014temperature}. When adopting the AHC formalism,\cite{allen1976theory,allen1981theory} in our calculations, we assume the on-mass-shell approximation i.e., $\omega = \varepsilon_{n\bfk}$ and the adiabatic approximation, i.e., $\varepsilon_{n\bfk}-\varepsilon_{m\bfk+\bfq} \gg \omega_{\bfq\nu}$. However, in some cases discussed below we did not adopt the adiabatic approximation.  Within the AHC formalism, the real part of the Fan-Migdal self-energy can be simplified:
\begin{equation}
    \mathbf{Re}\,\Sigma^{FM}_{n\bfk}(T) \approx \sum_{m\nu\bfq}|g_{mn\nu}(\bfk,\bfq)|^2\left[ \frac{2n_{\bfq\nu}+1}{\varepsilon_{n\bfk}-\varepsilon_{m\bfk+\bfq}} \right].
    \label{eq:FMAHC}
\end{equation}

To avoid summations over empty bands in the evaluation of Fan-Migdal and Debye-Waller self-energies (Eq.~\ref{eq:FM}-\ref{eq:FMAHC}), here we use the Lanczos approach. Writing  the Fan-Migdal self-energy within the AHC approximation (Eq.~\ref{eq:FMAHC}) as an example, we first expand the electron-phonon coupling matrix using Eq.~\ref{eq:g},
\begin{equation}
    \Sigma_{n\bfk}^{FM} = \sum_{m\nu\bfq}
    \left\langle\psi_{n\bfk}\left|\partial_{-\bfq\nu}V_{SCF} \right|\psi_{m\bfk+\bfq} \right\rangle \frac{2n_{\bfq\nu}+1}{\varepsilon_{n\bfk}-\varepsilon_{m\bfk+\bfq}}
    \left\langle \psi_{m\bfk+\bfq}\left|\partial_{\bfq\nu}V_{SCF}\right|\psi_{n\bfk} \right\rangle.
\end{equation}
Given the projection of the Hamiltonian on the conduction (virtual) manifold, i.e., $\tilde{H}_{\bfk+\bfq}=\hat{\mathcal{P}}_{\bfk+\bfq}^c H\hat{\mathcal{P}}^c_{\bfk+\bfq}$, we can write
\begin{equation}
    \sum_{m}\left|\psi_{m\bfk+\bfq}\right\rangle (\varepsilon_{n\bfk}-\varepsilon_{m\bfk+\bfq})^{-1}\left\langle\psi_{m\bfk+\bfq}\right| = (\varepsilon_{n\bfk}-\tilde{H}_{\bfk+\bfq})^{-1}.
    \label{eq:operator}
\end{equation}
Eq.~\ref{eq:operator} may be efficiently solved using the Lanczos approach (a detailed derivation is presented in the SI), and the imaginary part of the Fan-Midgal self energy and the Debye-Waller self-energy (Eq.~\ref{eq:FM}-\ref{eq:DW}) may be computed  in a similar manner. 

The approach described above was implemented in the \texttt{WEST} code\cite{govoni2015large,mcavoy2018coupling}, interfaced with \texttt{Quantum Espresso} (version 6.1)\cite{giannozzi2009quantum} and the symmetries at $\bfk$ and $\bfq$ points were analyzed using the \texttt{PHonon} package\cite{giannozzi2009quantum}.

A brief summary of the main features of the methodology presented here, compared to the implementations of first principles electron-phonon calculations used in the current literature, is given in Table \ref{tab:compare_codes}. 

\begin{table}[]
    \centering
    \begin{threeparttable}
    \begin{tabular}{p{0.4\textwidth}|ccc}
    \hline\hline
    & Ref.~\citenum{giustino2010electron} & Ref.~\citenum{antonius2014many} & This work \\
    \hline
    Frequency Integration & PPM & PPM & FF\\
    \hline
    Evaluation of $\Sigma_{G_0W_0}$ & S\tnote{a} & S & NS \\
    \hline
    Evaluation of $\Sigma_\mathrm{ep}$ & S & S/NS\tnote{b} & NS\\
    \hline
    Combined evaluation\\ 
    of $\Sigma_{G_0W_0}$ and $\Sigma_\mathrm{ep}$ & N & N & Y \\
    \hline\hline
    \end{tabular}
    \begin{tablenotes}
    \item[a] The $G_0W_0$ energy levels were not used in the evaluation of electron-phonon self-energies at $G_0W_0$ level, instead a scissor operator corresponding to the $G_0W_0$ correction was applied to DFT energy levels.
    \item[b] No empty bands were used in the evaluation of electron-phonon self-energies at the DFT level. No information was provided on the calculations of empty states for $G_0W_0$ calculations.
    \end{tablenotes}
    \end{threeparttable}
    \caption{First principles calculations of electron-phonon self-energies based on the $G_0W_0$ approximation. The integral  of the self energy as a function of frequency is evaluated using either a plasmon-pole model (PPM) or by carrying out  full-frequency (FF) integration using contour deformation.\cite{govoni2015large,godby1988self}  Evaluation of the $G_0W_0$ self-energy ($\Sigma_{G_0W_0}$) and of the electron-phonon self-energy ($\Sigma_\mathrm{ep}$) are performed with algorithms requiring summation (S) over virtual states or no summation (NS) over virtual states. The evaluation of $\Sigma_{G_0W_0}$ and $\Sigma_\mathrm{ep}$ is combined in this work (Y) but carried out separately (N) in previous works. }
    \label{tab:compare_codes}
\end{table}

\section{Verification protocol}\label{sec:verification}

Our verification protocol includes first the comparison of phonon frequencies computed with standard DFPT based approaches with those of our methodology (Eq.~\ref{equ:dVscf}); we then carry out a study of the numerical parameters affecting the calculations of the ZPR of diamond within DFT and $G_0W_0$ and compare our results with those present in the literature.

{\it Phonon frequencies} -- To verify our implementation, we first compared the phonon frequencies of the diamond crystal obtained with the method described above to those computed with the \texttt{PHonon} package in \texttt{Quantum Espresso}\cite{giannozzi2009quantum}. For verification purposes we carried out our calculations using the local density approximation (LDA), a cutoff of 60 Ry, Trouiller-Martins type pseudopotentials\cite{troullier1991efficient} generated with the \texttt{FHI98pp} code\cite{fhi98pp}, and a $3\times3\times3$ $\bfk$-point mesh. 
Fig.~S1 shows that the interpolated phonon dispersion curves in diamond obtained with the two approaches are indistinguishable, with a mean absolute difference less than  $1\,\mathrm{cm^{-1}}$. The comparison was repeated using energy cutoffs of 80, 100, and 120 Ry, and we found that the remaining small discrepancies not visible on the figure can be further reduced by increasing the cutoff (see Fig.~S2). 



{\it Zero point renormalization of energy levels at the DFT level of theory} -- 
We now turn to the discussion of electron-phonon self-energies. The real and imaginary parts of the  electron-phonon self-energy yield the zero point renormalization and the lifetime of the single particle energy levels, respectively. To verify our implementation, we computed the ZPR of single particle energy levels in diamond at the DFT/LDA level of theory and compared our results with those reported in Ref.~\citenum{antonius2014many} and Ref.~\citenum{ponce2014temperature}. In these two papers, the ZPRs are computed as 
\begin{equation}
    \Delta\varepsilon_{n\bfk}(T) = \frac{1}{N_\bfq}\sum_{\bfq\nu}\frac{\partial\varepsilon_{n\bfk}}{\partial n_{\bfq\nu}}\left[n_{\bfq\nu}(T)+\frac{1}{2}\right],\label{equ:EPCE}
\end{equation}
where $N_\bfq$ is the number of $\bfq$ points and $\partial\varepsilon/\partial n$ is the electron-phonon coupling energy (EPCE). The latter can be evaluated using frozen-phonon  or DFPT calculations,\cite{ponce2014temperature,monserrat2018electron} and here we report our results using DFPT  and the AHC approximation,\cite{allen1976theory,allen1981theory,ponce2014temperature}
\begin{equation}
    \frac{\partial \varepsilon^{FM}_{n\mathbf{k}}}{\partial n_{\nu\mathbf{q}}}= 2\sum_{m}\frac{\left|g_{nm\nu}(\mathbf{k},\mathbf{q})\right|^2}{\varepsilon_{n\mathbf{k}}-\varepsilon_{m\mathbf{k+q}}},
\end{equation}
\begin{equation}
\frac{\partial \varepsilon^{DW}_{n\mathbf{k}}}{\partial n_{\nu\mathbf{q}}} = -\frac{1}{2\omega_{\mathbf{q}\nu}}\sum_{m}\sum_{I\alpha J\beta}\frac{1}{\varepsilon_{n\mathbf{k}}-\varepsilon_{m\mathbf{k}}}\left[\frac{\xi_{I\alpha,\mathbf{q}\nu}\xi_{I\beta,\mathbf{q}\nu}^*}{M_I} +\frac{\xi_{J\alpha,\mathbf{q}\nu}\xi_{J\beta,\mathbf{q}\nu}^*}{M_J}\right]g^{*,I\alpha}_{mn}(\mathbf{k},\mathbf{0})g^{J\beta}_{mn}(\mathbf{k},\mathbf{0}).
\end{equation}
We computed EPCEs with the Troullier-Martins type pseudopotential\cite{troullier1991efficient}, an energy cutoff of 60 Ry, and $6\times 6\times 6$ $\bfk$-point sampling, as in Ref. \citenum{ponce2014temperature}. We performed two sets of calculations, one using the lattice constant that we optimized at the LDA level (3.5185 \AA), and the other using the lattice parameter (3.5323 \AA) reported in Ref.~\citenum{ponce2014temperature}. In \autoref{tab:epce_compare} we compare the computed EPCEs with those in Ref.\citenum{ponce2014temperature} at $\bfk = \Gamma,\,\,L$ and $\bfq = \Gamma,\,\, L$. We find a mean absolute difference less than $3\,\mathrm{meV}$ and the mean absolute relative difference is $\sim 2\,$\%. The largest differences are observed at $(\bfq=\Gamma,\,\bfk=L_3)$ and $(\bfq=L,\,\bfk=\Gamma_{2^\prime})$. For $(\bfq=\Gamma,\bfk=L_3)$, the EPCE computed with the optimized structure is $-162.66\,\mathrm{meV}$ and the value reported in Ref.\citenum{ponce2014temperature} is $-180.55\,\mathrm{meV}$; for $(\bfq=L,\bfk=\Gamma_{2^\prime})$, the EPCE computed with our code and the lattice constant of Ref.\citenum{ponce2014temperature} is $-294.70\,\mathrm{meV}$, to be compared to $-307.54\,\mathrm{meV}$, reported in Ref.\citenum{ponce2014temperature}. 

\begin{table}[htbp]
    \centering
    \begin{tabular}{cc|S[table-number-alignment = center]S[table-number-alignment = center]|S[table-number-alignment = center]}
    \hline\hline
    \textbf{$\bfq$ point} & \textbf{$\bfk$ point} & \textbf{Optimized Cell} & \textbf{Cell parameter from Ref.~\citenum{ponce2014temperature}} & \textbf{Ref. \citenum{ponce2014temperature}} \\
    \hline
    $\Gamma$ & $\Gamma_1$           &  -12.55 &   -12.84     &  -12.53 \\
             & $\Gamma_{25^\prime}$ &   25.13 &    24.86     &   24.83 \\
             & $\Gamma_{15}$        &  -14.87 &   -14.88     &  -14.23 \\
             & $\Gamma_{2^\prime}$  &  -31.91 &   -30.86     &  -30.93 \\
             & $L_{2^\prime}$       &  -21.18 &   -21.54     &  -20.98 \\
             & $L_1$                &  -16.72 &   -16.91     &  -16.60 \\
             & $L_{3^\prime}$       &   10.14 &    10.02     &   10.10 \\
             & $L_3$                & -162.66 &  -182.88     & -180.55 \\
             \hline
    $L$      & $\Gamma_1$           &  -54.47 &   -55.15     &  -53.73 \\
             & $\Gamma_{25^\prime}$ &  186.17 &   186.71     &  181.28 \\
             & $\Gamma_{15}$        & -273.16 &  -274.86     & -273.58 \\
             & $\Gamma_{2^\prime}$  & -311.89 &  -294.70     & -307.54 \\
             & $L_{2^\prime}$       &  -91.27 &   -91.96     &  -89.36 \\
             & $L_1$                & -212.64 &  -224.15     & -220.56 \\
             & $L_{3^\prime}$       &  -26.88 &   -27.13     &  -25.91 \\
             & $L_3$                &  163.96 &   164.07     &  163.19 \\
    \hline
    \multicolumn{2}{c}{MAD [meV]} & 2.63 & 2.10 &  \\
    \multicolumn{2}{c}{MARD [\%]} & 2.29\,\%  & 2.13\,\% & \\
    \hline\hline
    \end{tabular}
    \caption{Electron-phonon coupling energies (see Eq.~\ref{equ:EPCE}) [meV] computed with optimized cell parameters (third column) and with the cell parameters reported by Ref.~\citenum{ponce2014temperature} (fourth column). Mean absolute differences (MAD)[meV] and mean absolulte relative differences (MARD) are given in the last row.}
    \label{tab:epce_compare}
\end{table}

Ref. \citenum{ponce2014temperature} reported EPCEs' values but did not report renormalizations of energy levels in diamond. Therefore, to verify our computed renormalizations, we compare our results with those of Ref.~\citenum{antonius2014many}, using the same lattice parameter (3.5473 \AA) and $4\times 4\times 4$ $\bfk$-point sampling.  The computed renormalization of the highest occupied  and lowest unoccupied bands at the $\Gamma$ point are 116 and $-319\,\mathrm{meV}$, respectively,  in good agreement with the values of 113 and $-314\,\mathrm{meV}$, reported in Ref.\citenum{antonius2014many}. As a result, our computed renormalization of the direct gap in diamond is $-439\,\mathrm{meV}$, which also agrees very well with the result $-427\,\mathrm{meV}$ of  Ref. \citenum{antonius2014many}. When using the lattice constant (3.5185 \AA) optimized in our calculations, we obtain a renormalization energy  of $-429\,\mathrm{meV}$ and the difference compared to previously published values  is only $2\,\mathrm{meV}$.

We also evaluated the Fan-Migdal self-energy without adopting the AHC approximation and thus considering so called non-adiabatic terms, by including phonon frequencies in the denominator of Eq.~\ref{eq:FM}. We emphasize that in our approach, which  does not require summations over empty bands, the inclusion of non-adiabatic effects comes at no extra computational cost, as does  the evaluation of electron-phonon self energies at multiple temperatures and frequencies. We found that the  ZPR of the indirect band gap of diamond computed by including non-adiabatic effects is $-332\,\mathrm{meV}$, in good agreement with the value $-327\,\mathrm{meV}$ reported in Ref.~\citenum{ponce2014temperature}, where the LDA functional and $10\times10\times10$ $\bfk$-point sampling were used, as in our work.

\begin{figure}[htbp]
    \centering
    \begin{subfigure}[b]{0.8\textwidth}
    \includegraphics[width=\textwidth]{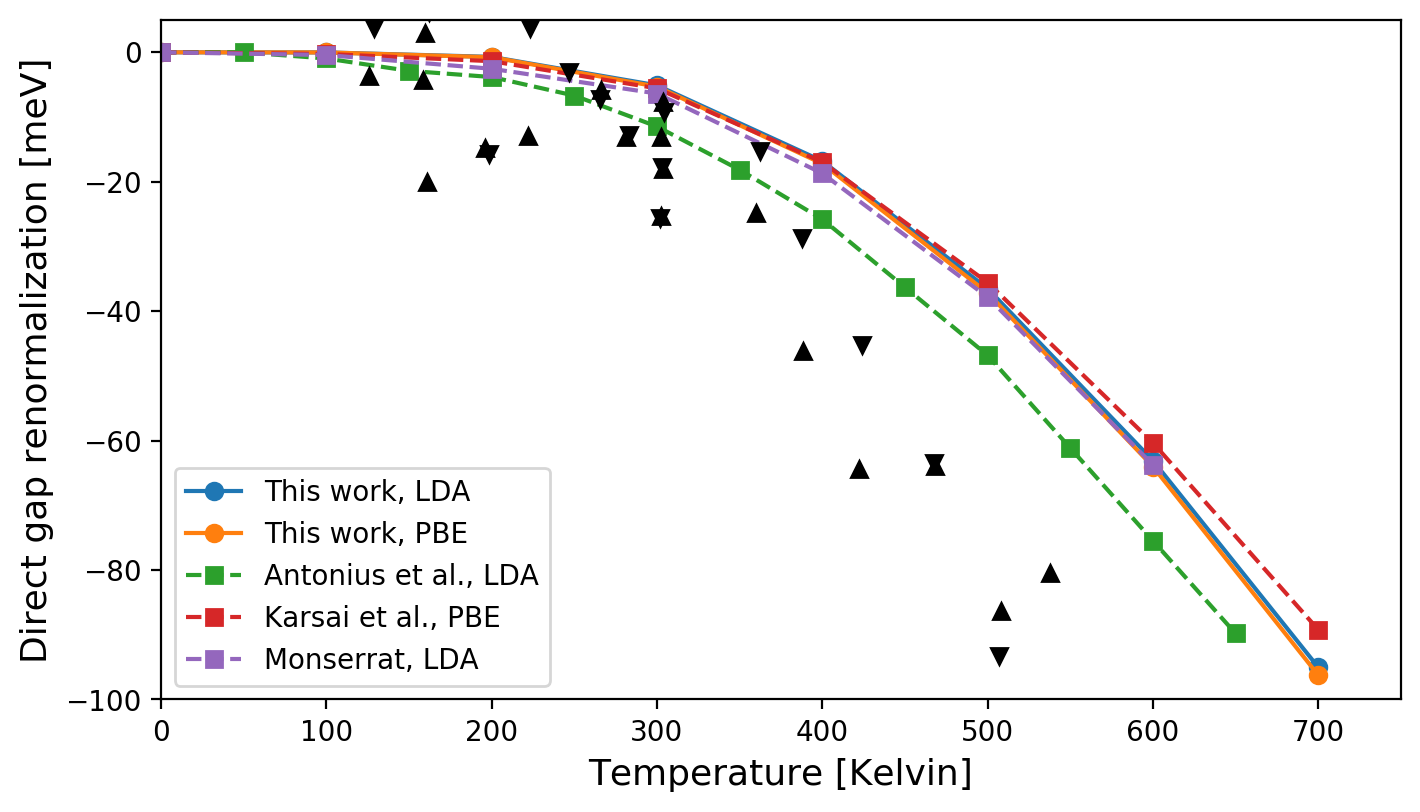}
    \end{subfigure}
        \begin{subfigure}[b]{0.8\textwidth}
    \includegraphics[width=\textwidth]{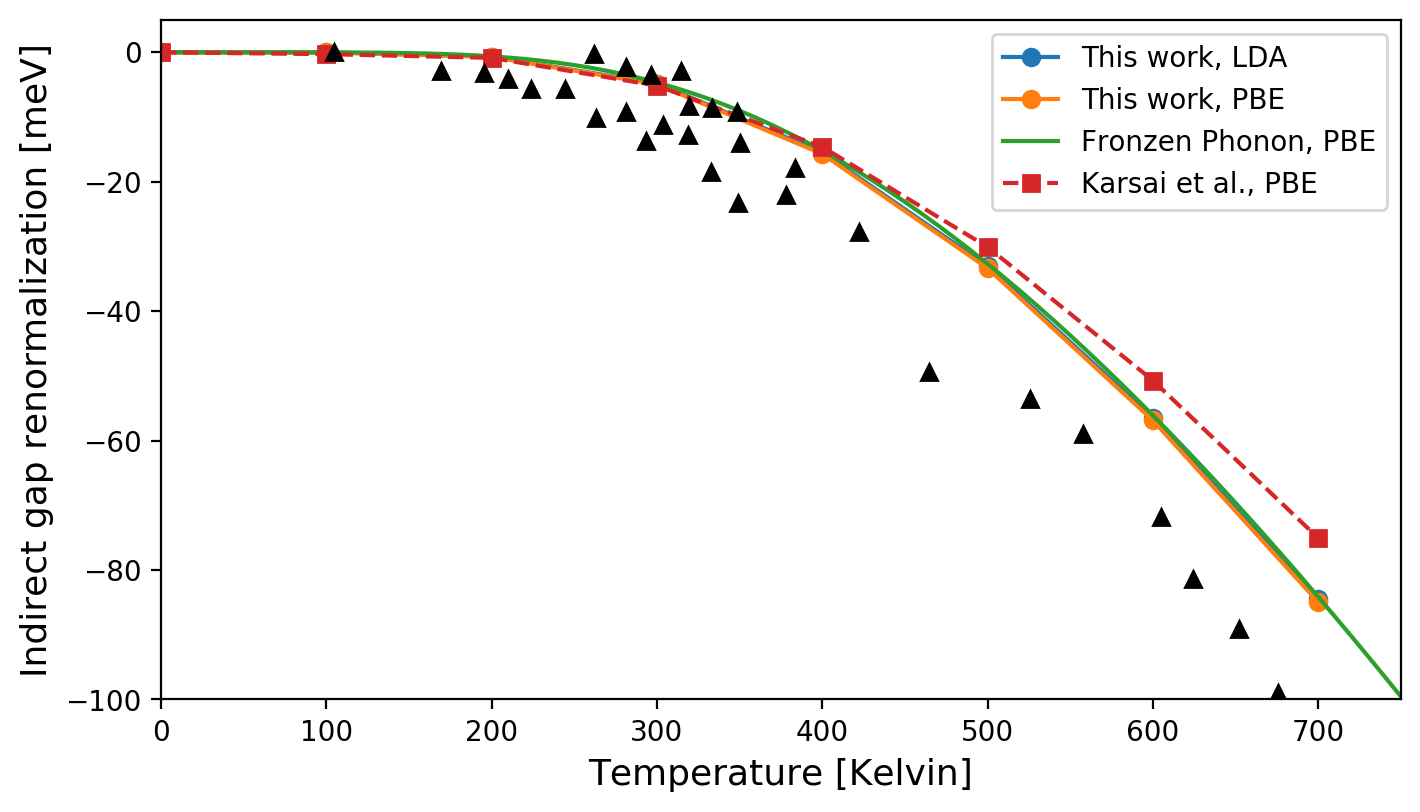}
    \end{subfigure}
    \caption[width=0.8\textwidth]{Temperature dependence of direct (upper) and indirect (lower) band gap in diamond. The renormalization at zero temperature was set at zero. The literature calculations are \citeauthor{antonius2014many} (Ref.~\citenum{antonius2014many}), \citeauthor{karsai2018electron} (Ref.~\citenum{karsai2018electron}) and \citeauthor{monserrat2016correlation} (Ref.~\citenum{monserrat2016correlation}), and the experimental renormalizations (black triangles) of direct and indirect gap are extracted from Ref.~\citenum{logothetidis1992origin} and Ref.~\citenum{chen1991temperature}, respectively.}
    \label{fig:temperature_dependence}
\end{figure}

We close the discussion of our DFT results by presenting temperature-dependent renormalizations of both the direct and indirect gaps in diamond obtained with $4\times 4\times 4$ $\bfq$-point sampling (Fig.~\ref{fig:temperature_dependence}). We carried out the calculations with the LDA functional at the LDA lattice constant obtained in Ref.\citenum{antonius2014many}, and with the PBE functional and the optimized lattice constant at the PBE level of theory. We find an almost identical temperature dependence with the two functionals. Our results for the direct gap renormalization compare well with those of  Ref.~\citenum{monserrat2016correlation} and \citenum{karsai2018electron}; however, they differ from those of Ref. \citenum{antonius2014many}. As for the indirect gap, our results agree well with those obtained from  frozen-phonon calculations\cite{kundu2021quantum} and with the findings of Ref.~\citenum{karsai2018electron} for temperatures lower than $500\,\mathrm{K}$, however we find larger differences at higher temperature.



{\it Zero point renormalization of energy levels at the $G_0W_0$ level of theory} -- We conducted a detailed study of the zero point renormalization of the direct gap of diamond at the level of $G_0W_0$, using different numerical  protocols, denoted P1 -- P5 in \autoref{tab:zpr_comparison}. A detailed explanation of the numerical protocols can be found in the SI. We note that in our electron-phonon calculations at the $G_0W_0$ level, we apply  $G_0W_0$ corrections only to DFT energy levels and we compute the electron-phonon coupling matrix elements at the DFT level. Our  fully-converged result is $-545\,\mathrm{meV}$ (P1 in \autoref{tab:zpr_comparison}), which is smaller than the value reported in Ref. \citenum{antonius2014many}.  We investigated the dependence of the results on $N_\mathrm{states}$, the number of states used in the evaluation of the $G_0W_0$ self-energy; on whether the Lanczos approach was used in the evaluation of $G_0W_0$ and electron-phonon self-energies; and on whether a full-frequency (FF) integration or the Hybersten-Louie plasmon-pole model (PPM) was used in the evaluation of the $G_0W_0$ self-energy. The $G_0W_0$ calculation with the PPM was carried out with the \texttt{ABINIT} package.\cite{gonze2020abinit,bruneval2006effect} In addition, we used the curvature technique proposed in Ref.~\citenum{nguyen2009efficient} in the calculation of the  exchange part of the $G_0W_0$ self-energy.

In \autoref{tab:zpr_comparison}, P1 and P2 both yield what we consider a converged value of the ZPR $-545\,\mathrm{meV}$, obtained by using  8 bands (4 valence bands and 4 conduction bands) and 100 bands (4 valence bands and 96 conduction bands), respectively.  In P3, the curvature correction was not adopted when computing the exchange part of the electron self-energies and the computed ZPR, $-562\,\mathrm{meV}$, was about 20 meV lower than our converged value. In P4, where the Lanczos approach was not adopted, we obtained an even lower value,  $-600\,\mathrm{meV}$. For P1 -- P4, the $G_0W_0$ quasiparticle energies were obtained with the \texttt{WEST} code\cite{govoni2015large} with full-frequency (FF) integration. Finally in P5 we used the $G_0W_0$ band structure obtained with the plasmon-pole model (PPM) computed with the \texttt{ABINIT} package, as input for our calculations. By doing so we obtain a ZPR of $-620\,\mathrm{meV}$, in good agreement with the result $-622\,\mathrm{meV}$ reported in  Ref.\citenum{antonius2014many}.

 This comparison shows that the accuracy of $G_0W_0$ corrections to DFT eigenvalues has a significant impact on the computed electron-phonon self-energies; the comparison also shows that the plasmon-pole model may not be sufficiently accurate  even for a simple crystal such as diamond. 

\begin{table}[htbp]
    \centering
    \begin{tabular}{c|cccccc}
    \hline\hline
                      & \textbf{P1} & \textbf{P2} & \textbf{P3} & \textbf{P4} & \textbf{P5} & \textbf{Ref. \citenum{antonius2014many}} \\
    \hline
    ZPR [meV]        &             -545 &   -545 &   -562 &   -600 &   -620 & -622 \\
     \hline
    $N_\mathrm{states}$ ($\Sigma_{G_0W_0}$) &               8 &    100 &    100 &    100 &    100 & 100 \\
    Lanczos ($\Sigma_{G_0W_0}$) &             Yes &    Yes &    Yes &     No &     No & No \\
    Lanczos ($\Sigma_\mathrm{ep}$)     &             Yes &    Yes &    Yes &    Yes &    Yes & See text \\
    Freq. integration &              FF &     FF &     FF &     FF &     PPM & PPM \\
    Curvature         &             Yes &    Yes &     No &     No &     No & N/A \\
    \hline\hline
    \end{tabular}
    \caption{Zero point renormalization (ZPR) of the direct gap of diamond obtained with different computational protocols (P) at the level of $G_0W_0$@LDA. $N_\mathrm{states}$ denotes the number of empty bands used in the evaluation of the $G_0W_0$ self-energy; Lanczos the algorithm used for the frequency integration; FF and PPM stand for full frequency and plasmon pole model, respectively. In the last row we indicate whether the curvature technique of Ref.~\citenum{nguyen2009efficient} was included in the calculation of the exact exchange term of the self-energy. See SI}
    \label{tab:zpr_comparison}
\end{table}

\section{Large scale calculations: zero point renormalization in defective solids} 
\begin{figure}
    \centering
    \begin{subfigure}[b]{0.45\textwidth}
    \includegraphics[width=\linewidth]{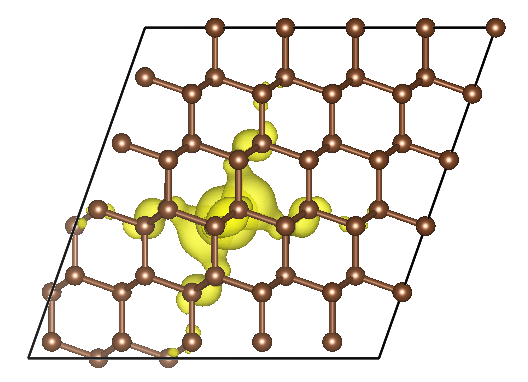}
    \end{subfigure}
    \begin{subfigure}[b]{0.45\textwidth}
    \includegraphics[width=\linewidth]{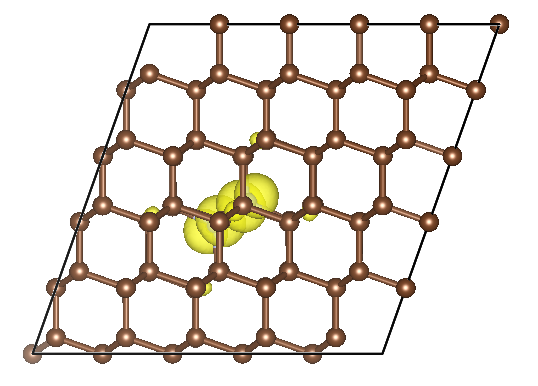}
    \end{subfigure}
    \caption{Isosurface (yellow) of the square moduli of the single particle orbitals of the boron pair unoccupied defect state (left) and nitrogen pair occupied defect state (right), as obtained in a $5\times 5\times 5$ supecell of diamond.}
    \label{fig:defect_structure}
\end{figure}

After verifying our implementation and examining the effect of various numerical approximations, we carried out calculations for supercells representative of defective solids, in particular defects in diamond (see \autoref{fig:defect_structure}), showcasing the ability of the methodology developed here to carry out calculations for large systems. We considered two nearest neighbor carbon atoms replaced by either two boron or two nitrogen atoms. The electronic structure of the boron (nitrogen) pair exhibits one unoccupied (occupied) state in the band gap of the host. We carried out DFT calculations with the PBE functional\cite{perdew1996PBE}, SG15\cite{schlipf2015SG15} ONCV\cite{hamann2013ONCV} pseudopotentials and an energy cutoff of 50 Ry. The $G_0W_0$ calculations were carried out with the \texttt{WEST} code. We report the electron-phonon renormalizations of the systems with defects in \autoref{tab:boron_defect_zpr} and \autoref{tab:nitrogen_defect_zpr}.

We first discuss the results using the AHC approximation (\autoref{tab:boron_defect_zpr} and \autoref{tab:nitrogen_defect_zpr}). The presence  of defects affects  the renormalizaton of the gap of the host crystal due to symmetry breaking,\cite{kundu2021quantum} with the effect decreasing in magnitude  as the size of supercell increases, as expected. In order to find a converged value of the renormalization, we extrapolated our computed values as a function of the inverse of the supercell size for defective systems, and as a function of the inverse of the number of $\bfq$-points for pristine diamond. Our extrapolated values at the PBE level are $-340\,\mathrm{meV}$ for the gap of pristine diamond and  $-338$ and $-370\,\mathrm{meV}$ for that of diamond with boron and nitrogen pairs, respectively. Once we add  $G_0W_0$ corrections, the corresponding extrapolated values are  $-351\,\mathrm{meV}$ and  $-372$ and $-386\,\mathrm{meV}$. In all cases, within the AHC approximations, we do not find significant differences between the renormalizations of the gap of pristine and defective diamond.  

The renormalization of the defect state within the band gap is more interesting, as the renormalizations of the defect states arising from boron and nitrogen pairs are noticeably distinct,  $51\,\mathrm{meV}$ and  $-7\,\mathrm{meV}$, respectively, at the PBE level, after extrapolation. When adding $G_0W_0$ corrections, the extrapolated values are $40\,\mathrm{meV}$ and $1\,\mathrm{meV}$. 



\begin{table}[]
    \centering
    \begin{threeparttable}
    \begin{tabular}{ccccc}
    \hline\hline
    & Supercell           & ZPR (Defect)  & ZPR (Host) & ZPR (Pristine) \\
    \hline
    \multirow{7}{*}{Adiabatic DFT/PBE} & $3\times 3\times 3$                 & 36       & -397                    & -277 \\
    & $4\times 4\times 4$                 & 46 & -401                           & -366 \\
    & $5\times 5\times 5$                 & 47 & -327                           & -316 \\
    & $6\times 6\times 6$  & --                  & -- & -324 \\
    & $8\times 8\times 8$ & --                  & -- & -330 \\
    & $10\times 10\times 10$  & --               & -- & -341 \\
    & extrapolated  & 51\tnote{a} & -338\tnote{a} & -357\tnote{a}, -340\tnote{b} \\
    \hline
    \multirow{4}{*}{Adiabatic $G_0W_0$@PBE}& $3\times 3\times 3$                 & 29  & -416                          & -254 \\
    & $4\times 4\times 4$                 & 37 & -448                          & -380 \\
    & $5\times 5\times 5$              & 37 & -346                            & -291 \\
    & extrapolated   & 40\tnote{a} & -372\tnote{a} & -351\tnote{a} \\
    \hline
    \multirow{7}{*}{Non-adiabatic DFT/PBE} & $3\times 3\times 3$                &    -267  & -556                       & -637  \\
    & $4\times 4\times 4$                & -152 &  -482                         & -510 \\
    & $5\times 5\times 5$                 & -131 & -362                          &  -372 \\
    & $6\times 6\times 6$ & --                  & -- & -353 \\
    & $8\times 8\times 8$ & --                  & -- & -330 \\
    & $10\times 10\times 10$ & --               & -- & -329 \\
    & extrapolated & -86\tnote{a} & -346\tnote{a} & -338\tnote{a}, -320\tnote{b} \\
    \hline
    \multirow{4}{*}{Non-adiabatic $G_0W_0$@PBE}& $3\times 3\times 3$ &	-275 & -574                             &  -615 \\
    & $4\times 4\times 4$  &	-163 & -518                           &  -508 \\
    & $5\times 5\times 5$  & -143 & -379                            &  -348 \\
    & extrapolated & -99\tnote{a} & -373\tnote{a} & -325\tnote{a} \\
    \hline\hline
    \end{tabular}
    \begin{tablenotes}
    \item[a] Extrapolated up to $5\times 5\times 5$
    \item[b] Extrapolated up to $10\times 10\times 10$
    \end{tablenotes}
    \end{threeparttable}
    \caption{Zero point renormalization (ZPR) [meV] of the energy gap of pristine diamond evaluated with k-point sampling (last column), of the energy gap of a supercell of diamond hosting a boron defect (Host), and of the state of a boron impurity in a supercell of diamond (Defect).}
    \label{tab:boron_defect_zpr}
\end{table}

\begin{table}[]
    \centering
    \begin{threeparttable}
    \begin{tabular}{ccccc}
    \hline\hline
    & Supercell            & ZPR (Defect) & ZPR (Host) & ZPR (Pristine) \\
    \hline
    \multirow{7}{*}{Adiabatic DFT/PBE} & $3\times 3\times 3$                 & 22   & -322                       & -277 \\
    & $4\times 4\times 4$                & 3  & -394                           & -366 \\
    & $5\times 5\times 5$                & 1   & -333                         & -316 \\
    & $6\times 6\times 6$ & --                  & -- & -324 \\
    & $8\times 8\times 8$ & --                  & -- & -330 \\
    & $10\times 10\times 10$ & --               & -- & -341 \\
    & extrapolated & -7\tnote{a} & -370\tnote{a} & -357\tnote{a}, -340\tnote{b} \\
    \hline
    \multirow{4}{*}{Adiabatic $G_0W_0$@PBE} & $3\times 3\times 3$                 & 33     & -339                      & -254 \\
    & $4\times 4\times 4$                 & 12   & -415                        & -380 \\
    & $5\times 5\times 5$               & 9    & -346                         & -291 \\
    & extrapolated & 1\tnote{a}  & -386\tnote{a} & -351\tnote{a} \\
    \hline
    \multirow{7}{*}{Non-adiabatic DFT/PBE} & $3\times 3\times 3$  &	185 &                 -459                          & -637\\
    & $4\times 4\times 4$ 	& 156 &  -431                          &  -510 \\
    & $5\times 5\times 5$ &	149 &  -345                           &  -372 \\
    & $6\times 6\times 6$ & --                  & -- & -353 \\
    & $8\times 8\times 8$ & --                  & -- & -330 \\
    & $10\times 10\times 10$ & --               & -- & -329 \\
    & extrapolated  & 138\tnote{a} & -344\tnote{a} & -338\tnote{a}, -320\tnote{b} \\
    \hline
    \multirow{4}{*}{Non-adiabatic $G_0W_0$@PBE} & $3\times 3\times 3$  & 	192  & -473                                        &  -615 \\
    & $4\times 4\times 4$  &	161 &   -452                                          &  -508 \\
    & $5\times 5\times 5$ 	& 153  &    -359                          &  -348 \\
    & extrapolated  & 141\tnote{a} & -362\tnote{a} & -325\tnote{a} \\
    \hline\hline
    \end{tabular}
    \begin{tablenotes}
    \item[a] Extrapolated up to $5\times 5\times 5$
    \item[b] Extrapolated up to $10\times 10\times 10$
    \end{tablenotes}
    \end{threeparttable}
\caption{Zero point renormalization (ZPR) [meV] of the energy gap of pristine diamond evaluated with k-point sampling (last column), of the energy gap of a supercell of diamond hosting a nitrogen defect (Host), and of the state of a nitrogen impurity in a supercell of diamond (Defect).}
    \label{tab:nitrogen_defect_zpr}
\end{table}

We now turn to discuss non-adiabatic effects. We found that including non-adiabatic effects in Eq.~\ref{eq:FM} changes substantially the renormalizations computed for defect states in diamond, although it has a smaller effect on the gap of diamond. Our results are reported in \autoref{tab:boron_defect_zpr} and \autoref{tab:nitrogen_defect_zpr}. The gap of diamond varies by $-56$, $-35$ and $-12\,\mathrm{meV}$ when including non-adiabatic effect, in the case of the pristine solid, boron and nitrogen defective systems, respectively, at the PBE level in the $5\times 5\times 5$ supercell. The magnitude of the renormalization of defect states increases by approximately a factor of 3 for boron pairs and by more than a factor of 10 for nitrogen in $5\times5\times5$ supercell, when including non-adiabatic effects. After extrapolation, the renormalization of the boron defect state is $-86\,\mathrm{meV}$ (PBE) and $-99\,\mathrm{meV}$ ($G_0W_0$) and that of the nitrogen pair is $138\,\mathrm{meV}$ (PBE) and $141\,\mathrm{meV}$  ($G_0W_0$).

\begin{figure}
    \centering
    \includegraphics[width=\linewidth]{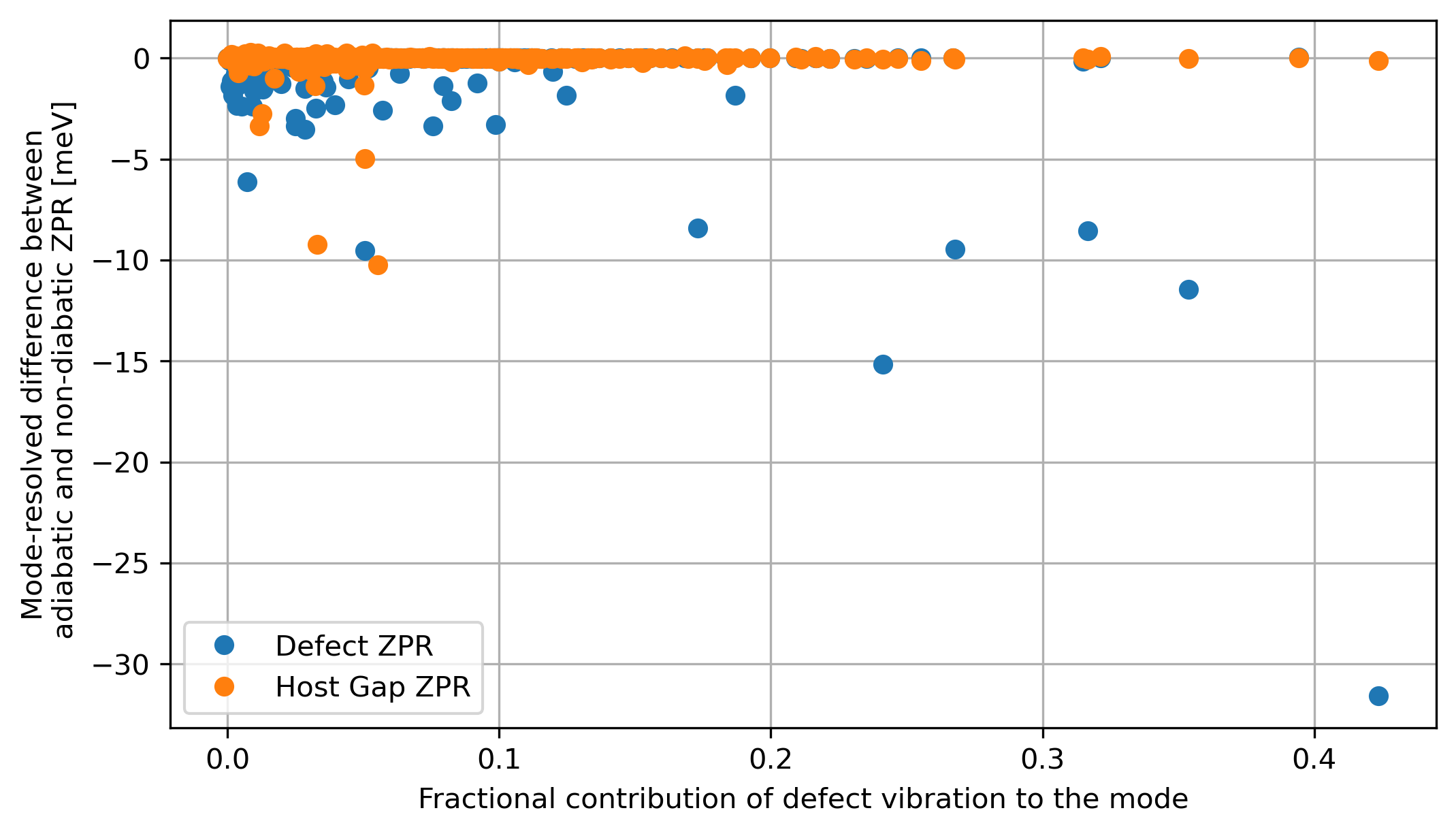}
    \caption{The contribution of vibration modes to the difference between adiabatic and non-adiabatic renormalizations for a boron pair in diamond, described in a $5\times5\times5$ supercell, with respect to the fractional contribution of the vibration of defect atoms $f_{\bfq\nu}^\mathrm{defect}$ (see text).}
    \label{fig:defect_mode_analysis}
\end{figure}

To understand the difference between the results obtained with and without non-adiabatic effects, we plot in \autoref{fig:defect_mode_analysis} the contribution of each vibrational mode to the difference between the adiabatic and non-adiabatic renormalizations for the boron pair, described with a  $5\times5\times5$ supercell; such difference is expressed in terms of the fractional contribution of defect atoms to each mode, $f^{\text{defect}}_{\bfq\nu}=\sum_{I}^\mathrm{defect}\sum_{\alpha}|\xi_{I\alpha,\bfq\nu}|^2$, where the summation runs over the boron atoms.
We define vibrational modes with $f_{\bfq\nu}^\mathrm{defect}>10\%$ as modes exhibiting defect relevant vibrations. \autoref{fig:defect_mode_analysis} shows that defect relevant vibrations are indeed responsible for the difference between adiabatic and non-adiabatic effects found in the case of defect states; however their  contribution to the host gap renormalization are small. Quantitatively, the defect relevant vibrations contribute  approximately $-89\,\mathrm{meV}$ to the difference between adiabatic and non-adiabatic defect state renormalizations, with the remaining  $-88\,\mathrm{meV}$ being accounted for by coupling with lattice vibrations. Overall our results indicate that the AHC formalism and related adiabatic approximation are  not sufficiently accurate to describe the electron-phonon renormalizations of carbon-based defect states, and that taking into account non-adiabatic effects is critical to obtain accurate results.

\section{Conclusions}
In summary, we presented an efficient, combined approach to compute  electron-electron and electron-phonon self-energies in solids, which can be used for large scale calculations, and enables the inclusion of non-adiabatic and temperature effects in a simple, straightforward manner, at no extra computational cost. 
This approach is a generalization to solid of the method proposed for molecules in Ref.~\citenum{mcavoy2018coupling}.  We discussed in detail verification and validation strategies for calculations at the DFT and $G_0W_0$ level of theory; we found  that the numerical accuracy of $G_0W_0$ band structures is critical to obtain robust predictions of zero point renormalizations of energy levels and that carrying out full frequency integration is necessary to reach the required accuracy. We presented calculations for pristine diamond and defects in diamonds with supercells containing $\sim 1000$ electrons; we found that while the inclusion of non-adiabatic effects leads to moderate changes in the renormalization of the diamond band gap, it is essential to obtain accurate results for defect levels in the gap. Work is in progress to apply our methodology to the study of spin-defects in diamond and in other insulators and semiconductors, including spin-phonon interaction.

\section{Associated content}
In the supporting information, we present a detailed derivation of the Lanczos approach. We also attached the pseudopontentials used in this work.

\begin{acknowledgement}
We thank He Ma and Ryan L. McAvoy for useful discussion. This work was supported by the Midwest Integrated Center for Computational Materials (MICCoM) as part of the Computational Materials Sciences Program funded by the U.S. Department of Energy. This research used resources of the National Energy Research Scientific Computing Center (NERSC), a DOE Office of Science User Facility supported by the Office of Science of the U.S. Department of Energy under Contract No. DE-AC02-05CH11231, resources of the Argonne Leadership Computing Facility, which is a DOE Office of Science User Facility supported under Contract DE-AC02-06CH11357, and resources of the University of Chicago Research Computing Center.
\end{acknowledgement}
\bibliography{ms.bib}

\clearpage


\end{document}


\maketitle

\section{Evaluation of electron-phonon self-energies with the Lanczos approach}

We discuss an approach based on the Lanczos method\cite{lanczos1950iteration}, to avoid summation over empty bands when computing the electron-phonon self-energy.
Staring from Eq.~17 of the main text, we define $A_{n\mathbf{k}}(\tilde{H}_{\mathbf{k}+\mathbf{q}}) = (\varepsilon_{n\bfk}-\tilde{H}_{\bfk+\bfq})^{-1}$, and Eq.~17 can be written as,
\begin{equation}
    A_{n\bfk}(\tilde{H}_{\bfk+\bfq}) = \sum_{m}\left|\psi_{m\bfk+\bfq}\right\rangle A_{n\bfk}(\varepsilon_{m\bfk+\bfq})\left\langle\psi_{m\bfk+\bfq}\right|
\end{equation}

Following references\cite{lanczos1950iteration,mcavoy2018coupling}, we obtain the Lanczos basis $\tilde{q}_l$ and corresponding eigenvalues $d_l$ of $\tilde{H}_{\bfk+\bfq}$, and thus the self-energy can be written as
\begin{equation}
    \Sigma_{n\bfk}^{FM}(T) = \sum_{\nu\bfq l}\left\langle L_{n\bfk}^{\bfq\nu} |\tilde{q}_{l}\right\rangle A_{n\bfk}(d_l)\left\langle \tilde{q}_l | R_{n\bfk}^{\bfq\nu} \right\rangle[2n_{\bfq\nu}(T)+1],\label{equ:FM_Lanczos}
\end{equation}
where $\left| L_{n\bfk}^{\bfq\nu} \right\rangle$ and $\left| R_{n\bfk}^{\bfq\nu} \right\rangle$ are vectors within the set $\{\left|\partial_{\bfq\nu}V_{SCF}\psi_{n\bfk}\right\rangle,\,n = 1,\,2,\,\cdots\}$.

In the following, we describe how to compute Lanczos basis functions, and hereafter we drop the superscripts and subscripts of $\left|L\right\rangle$ and $\left|R\right\rangle$ for simplicity. The Lanczos basis functions and eigenvalues can be obtained by diagonalizing the matrix
\begin{equation}
    Q^\dagger\tilde{H}Q = \left(
    \begin{matrix}
    \alpha_1 &  \beta_2 &          &         &          \\
    \beta_2  & \alpha_2 &  \beta_3 &         &          \\
             & \beta_3  & \ddots   & \ddots  &          \\
             &          & \ddots   & \ddots  & \beta_n  \\
             &          &          & \beta_n & \alpha_n \\
    \end{matrix}
    \right)
    \label{equ:Lanczos_matrix}
\end{equation}
where $Q = \{\left|q_l\right\rangle,\,l\,=\,1,\,2,\,\cdots,\,N_\mathrm{Lanczos}\}$ with $\left|q_1\right\rangle = \left|R\right\rangle$ are a set of orthonomal vectors, and the elements of the matrix are obtained from
\begin{equation}
    \alpha_n = \left\langle q_l \left| \tilde{H} \right| q_l\right\rangle
\end{equation}
and
\begin{equation}
    \beta_{n+1} = ||(\tilde{H}-\alpha_n)\left|q_n\right\rangle-\beta_n\left|q_{n-1}\right\rangle|| .
\end{equation}
The vectors  $\left|q_l \right\rangle$ are orthogonalized  with a recursive process by applying
\begin{equation}
    \left|q_{n+1}\right\rangle = \frac{1}{\beta_{n+1}}\left[(\tilde{H}-\alpha_n)\left|q_n\right\rangle-\beta_n\left|q_{n-1}\right\rangle\right].
\end{equation}
The diagonalization of Eq.~\ref{equ:Lanczos_matrix} yields the eigenvalues $d_l$ and corresponding eigenvectors $U_l$. We then define a modified basis set $\left| \tilde{q}_l \right\rangle$ as a linear combination of the original basis $\left| q_l \right\rangle$,
\begin{equation}
    \left| \tilde{q}_l \right\rangle  = \sum_{k}^{N_\mathrm{Lanczos}} U_l^k\left|q_k \right\rangle.
\end{equation}
Having obtained the eigenvalues $d_l$ of the matrix $Q^\dagger\tilde{H}Q$ and using the modified basis $\left| \tilde{q}_l \right\rangle$, we can evaluate the Fan-Migdal self-energy in Eq.~\ref{equ:FM_Lanczos}, without summations over empty bands. A similar technique can  be applied to obtain the Debye-Waller self-energy.

In addition, we can compute the temperature-dependent, non-adiabatic or frequency-dependent self-energies without any extra computational cost, by reusing the Lanczos basis set defined above.

\section{Convergence study of phonon frequencies in diamond}

\begin{figure}
    \centering
    \includegraphics[width=0.8\textwidth]{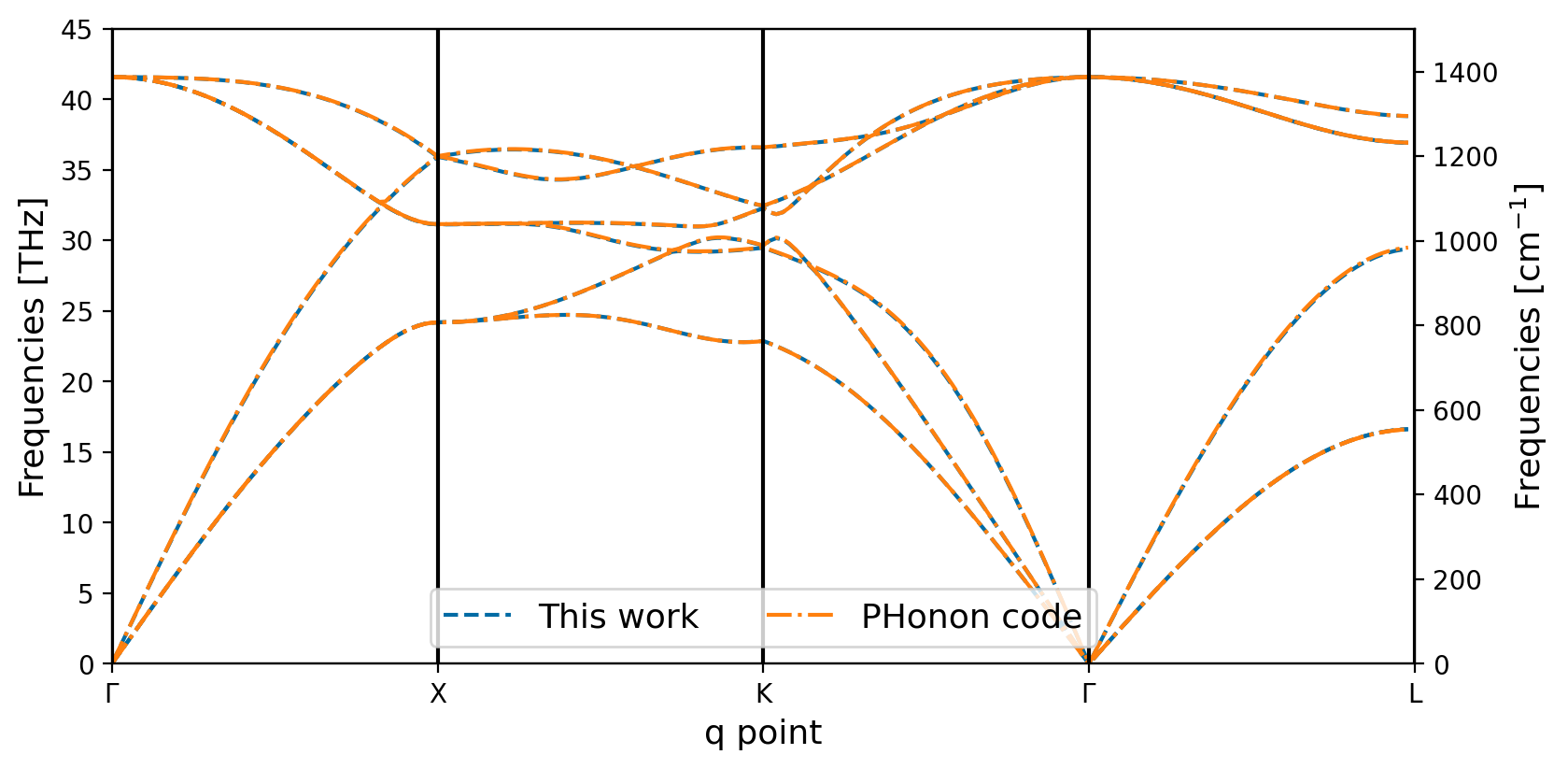}
    \caption{Phonon dispersion of diamond interpolated from $3\times3\times3$ $\bfq$-point sampling. A kinetic energy cutoff of 60 Ry and LDA functional were used. }
    \label{fig:diamond_dispersion}
\end{figure}

To verify our implementation, we computed phonon frequencies with the method described in the main text with the LDA functional and 60 Ry cutoff, and compared them with the results obtained with the \texttt{PHonon}\cite{giannozzi2009quantum} code.
Fig.~\ref{fig:diamond_dispersion} shows the phonon dispersion curves computed with our code and the \texttt{PHonon} code, which are indistinguishable.

\begin{figure}
    \centering
    \includegraphics[width=0.8\textwidth]{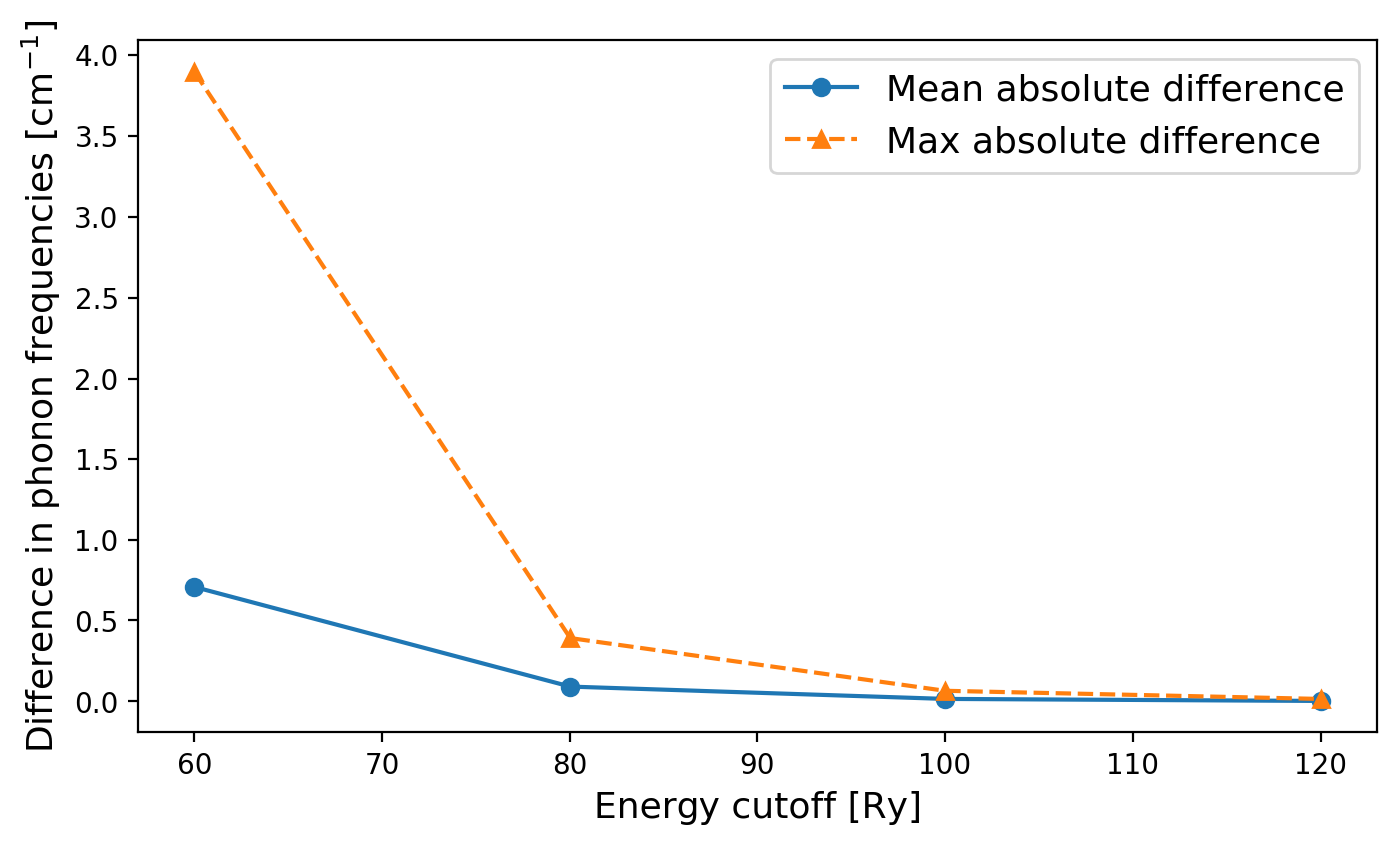}
    \caption{Mean and maximum absolute difference of phonon frequencies [in $\mathrm{cm}^{-1}$] computed with the method discussed in this work and the \texttt{PHonon} package in \texttt{Quantum Espresso}.}
    \label{fig:phonon_differences}
\end{figure}

We also carried out additional  comparisons with higher energy cutoffs and in Fig.~\ref{fig:phonon_differences} we report the differences of the phonon frequencies with different energy cutoffs. As the energy cutoff increases, both methods converge to the same phonon frequencies. 

\clearpage

\section{ Numerical protocol used in Table 3 of the main text}
The $G_0W_0$ self-energy $\Sigma$ contains an exchange term 
\begin{equation}
    \Sigma_x(\bfr,\bfr^\prime) = -\sum_{n=1}^{N_\mathrm{occ}}\sum_{\bfk}\psi_{n\bfk}(\bfr)v_c(\bfr,\bfr^\prime)\psi_{n\bfk}^*(\bfr^\prime)
\end{equation}
and a correlation term
\begin{equation}
    \Sigma_c(\bfr,\bfr^\prime;\omega) = i\int_{-\infty}^{+\infty}\frac{\mathrm{d}\omega^\prime}{2\pi}G_\mathrm{KS}(\bfr,\bfr^\prime;\omega+\omega^\prime)W_p(\bfr,\bfr^\prime;\omega^\prime),
\end{equation}  
where $\psi_{n\bfk}$ are Kohn-Sham orbitals associated with the $n$-th level at the $\bfk$ point, $v_c$ is the bare Coulomb potential, $G_\mathrm{KS}$ is the Green function written in terms of Kohn-Sham orbitals,
\begin{equation}
    G_\mathrm{KS}(\bfr,\bfr^\prime,\omega) = \sum_{n\bfk}\frac{\psi_{n\bfk}(\bfr)\psi^*_{n\bfk}(\bfr^\prime)}{\omega-\varepsilon_{n\bfk}}
\end{equation}
with $\varepsilon_{n\bfk}$ being the $n$-th Kohn-Sham energy at the $\bfk$ point, and $W_p$ is the difference between the screened Coulomb potential and bare Coulomb potential,
\begin{equation}
    W_p(\bfr,\bfr^\prime;\omega) = W(\bfr,\bfr^\prime;\omega) - v_c(\bfr,\bfr^\prime)
\end{equation}

To improve the convergence of the calculations of the exchange part $\Sigma_x$ with respect to the number of $\bfk$ points, the curvature technique developed by  Gygi-Baldereschi\cite{Gygi1986} and further refined by   Ref.~\citenum{nguyen2009efficient} was used in most of our calculations. In Table 3 of the main text, the curvature technique is used to obtain the results presented in P1 and P2, but not those of P3 -- P5.

The screened Coulomb interaction $W$ is evaluated by computing  the dielectric matrix $\epsilon$,
\begin{equation}
    W(\bfr,\bfr^\prime;\omega) = \epsilon^{-1}(\bfr,\bfr^\prime;\omega)v_c(\bfr,\bfr^\prime).
\end{equation}
and the symmetrized dielectric matrix $\tilde{\epsilon}$ is computed from the symmetrized polarizibility $\tilde{\chi}^0$,
\begin{equation}
    \tilde{\epsilon}_{\bfG\bfG^\prime}(\bfq,\omega) = \delta_{\bfG\bfG^\prime} - \tilde{\chi}^0_{\bfG\bfG^\prime}(\bfq,\omega).
\end{equation}
The symmetrized polarizibility can be written as:
\begin{equation}
\begin{aligned}
    \tilde{\chi}^0_{\bfG\bfG^\prime}(\bfq;\omega) = & -4\pi e^2\sum_n^{N_\mathrm{occ}}\sum_{m=N_\mathrm{occ}+1}^{+\infty}\sum_\bfk  \frac{\rho^*_{mn\bfk}(\bfq,\bfG)\rho_{jmn\bfk}(\bfq,\bfG^\prime)}{|\bfq+\bfG||\bfq+\bfG^\prime|} \\
    & \times\left[ \frac{1}{\varepsilon_{m\bfk}-\varepsilon_{n\bfk-\bfq}+\omega-i0^+} + \frac{1}{\varepsilon_{m\bfk}-\varepsilon_{n\bfk-\bfq}-\omega-i0^+}
    \right]
\end{aligned}
\end{equation}
with
\begin{equation}
    \rho_{mn\bfk}(\bfq,\bfG) = \left\langle \psi_{m\bfk} \left| \mathbf{e}^{i(\bfq+\bfG)\cdot\bfr}\right|\psi_{n\bfk-\bfq}\right\rangle
\end{equation}

The straightforward evaluation of the polarizibility $\tilde{\chi}^0$ is expensive because it requires the summation over empty bands and it is frequency dependent. In Table 3 of the main text, the calculations presented in the last column used 100 states for  the summation over empty bands. To compare our results with those if the literature, we also used 100 states for the results given in columns P2 -- P5. By using the Lanczos algorithm, we can avoid the summation over empty bands and there is no need to truncate the summation. For the results of the P1 column, only 8 bands are used, and we show  that the Lanczos algorithm yields the same result as that of the P2 column, where 100 bands are used.

In Table 3 of the main text, the calculation shown in the last column used the Plasmon-Pole model (PPM),\cite{ppm1986} a semi-empirical model, to compute the frequency dependence of the dielectric matrix, but our $G_0W_0$ calculation computes the full frequency (FF) dependence using the Lanczos approach without using any semi-empirical approximations. To compare our results with those existing in the literature, we used the PPM in column P5 and we did reproduce the literature result. However, FF is known to be more accurate than the PPM,\cite{golze2019gw} thus we used FF in obtaining the results of P1 -- P4.

The calculation of electron-phonon self-eneriges also requires to carry out  summations over empty bands, and we used the Lanczos technique for the electron-phonon self-energies in column P1 -- P5.

More details on the implementation of the $G_0W_0$ approximation can be found in Ref.~\citenum{govoni2015large}.

\section{Pseudopotentials used in this paper}
In \autoref{tab:Pseudopotentials}, we list the pseudopotentials that were used in this work.
\begin{table}[]
    \centering
    \begin{tabularx}{\textwidth}{lX}
    \hline\hline
    Pseudopotential & Comment  \\
    \hline
    C.pz-mt.fhi.UPF  & Martins-Troullier\cite{troullier1991efficient} type pseudopotential of carbon generated with Perdew-Zunger\cite{perdew1981self} parameterized LDA functional in UPF format and is ready for \texttt{Quantum Espresso} calculation. \\
    \hline
    C.pz-mt.fhi & Martins-Troullier\cite{troullier1991efficient} type pseudopotential of carbon generated with Perdew-Zunger\cite{perdew1981self} parameterized LDA functional in fhi format and is ready for \texttt{ABINIT} calculation. \\
    \hline
    C\_ONCV\_PBE-1.0.upf & SG15\cite{schlipf2015SG15} ONCV\cite{hamann2013ONCV} pseudopotential of carbon generated with PBE functional\cite{perdew1996PBE} and is ready for \texttt{Quantum Espresso} calculation. \\
    \hline
    B\_ONCV\_PBE-1.0.upf & SG15\cite{schlipf2015SG15} ONCV\cite{hamann2013ONCV} pseudopotential of boron generated with PBE functional\cite{perdew1996PBE} and is ready for \texttt{Quantum Espresso} calculation.\\
    \hline
    N\_ONCV\_PBE-1.0.upf & SG15\cite{schlipf2015SG15} ONCV\cite{hamann2013ONCV} pseudopotential of nitrogen generated with PBE functional\cite{perdew1996PBE} and is ready for \texttt{Quantum Espresso} calculation.\\
    \hline\hline
    \end{tabularx}
    \caption{Pseudopotentials used in this paper}
    \label{tab:Pseudopotentials}
\end{table}
\clearpage
\bibliography{supplement.bib}